# Ceteris Paribus Structure in Logics of Game Forms

Davide Grossi
Department of Computer Science
University of Liverpool

Emiliano Lorini
IRIT-CNRS, Université Paul Sabatier
Toulouse, France

François Schwarzentruber
ENS Cachan - Brittany extension - IRISA

## ABSTRACT

The article introduces a ceteris paribus modal logic interpreted on the equivalence classes induced by sets of propositional atoms. This logic is used to embed two logics of agency and games, namely atemporal STIT and the coalition logic of propositional control (CL−PC). The embeddings highlight a common ceteris paribus structure underpinning the key modal operators of both logics, they clarify the relationship between STIT and CL−PC, and enable the transfer of complexity results to the ceteris paribus logic.

## 1. INTRODUCTION

In a strategic game, the $\alpha$-effectivity of a set of players consists in those sets of outcomes of the game for which the players have some collective action which forces the game to end up in that set, no matter what the other players do [MP82]. So, if a set of outcomes $X$ belongs to the $\alpha$-effectivity of a set of players $J$, then each agent in $J$ can fix an individual action such that, for all actions of the other players, the game will end up in $X$.

It was already observed in [vBGR09] that the sort of reasoning underlying the notion of $\alpha$-effectivity is of a ceteris paribus nature. Evaluating the outcomes that can be reached in a game once a set of players $J$ has fixed their actions, amounts to considering what necessarily will be the case under the ceteris paribus condition 'all current actions of $J$ being equal'. It has been shown in [vBGR09] how this intuition can be used, for instance, to give a modal formulation of Nash equilibria.

The present paper builds on that idea and systematically explores the ceteris paribus structure of two main logics of agency and games based on the $\alpha$-effectivity concept: STIT [BPX01, Hor01] (the logic of *seeing to it that*) in its atemporal version [HS08], and the coalitional logic of propositional control (CL−PC) [vdHW05]. To articulate the analysis, whose main tool will consist of embedding results, the paper introduces and studies a simple ceteris paribus logic based on propositional equivalence.

*Structure of the paper.*
Section 2 introduces a logic called *propositional equivalence ceteris paribus logic* (PECP in short), which will be used as yardstick to analyze the game logics addressed in the paper. The logic will be axiomatized and briefly compared with existing modal logics of *ceteris paribus* reasoning.

Section 3 provides a study of the relationship between the atemporal version of STIT and PECP. We show that PECP embeds atemporal group STIT—the fragment of atemporal STIT in which both actions of individuals and groups are represented—under the assumption that the agents' choices are bounded. We call the latter atemporal 'bounded' group STIT. Moreover, we show that PECP embeds atemporal individual STIT—the variant of atemporal STIT in which only the actions of individuals are represented. The former embedding is used to transfer complexity results to PECP. We also present an embedding in PECP of a variant of atemporal group STIT in which groups are nested (i.e., given two sets of agents $J$ and $J'$ either $J \subseteq J'$ or viceversa).

Section 4 provides an embedding of coalition logic of propositional control into atemporal 'bounded' group STIT and, indirectly, it provides an embedding of coalition logic of propositional control into PECP.

We conclude in Section 5. Longer proofs are collected in a technical appendix at the end of the paper.

## 2. A CETERIS PARIBUS LOGIC BASED ON PROPOSITIONAL EQUIVALENCE

### 2.1 Equivalence modulo a set of atoms

Consider a structure $(W, V)$ where $W$ is a set of states, and $V : \mathbf{P} \longrightarrow 2^W$ a valuation function from a countable set of atomic propositions $\mathbf{P}$ to subsets of $W$. We define a simple notion of propositional equivalence between states in $W$, modulo subsets of $\mathbf{P}$.

DEFINITION 1. *(Equivalence modulo $X$) Given a pair $(W, V)$, $X \subseteq \mathbf{P}$ and $|X| < \omega$, the relation $\sim_X \subseteq W^2$ is defined as:*

$$w \sim_X^V w' \iff \forall p \in X : (w \in V(p) \iff w' \in V(p))$$

When $X$ is a singleton (e.g. $p$), we will often write $\sim_p^V$ instead of $\sim_{\{p\}}^V$. Also, in order to avoid clutter, we will often drop the reference to $V$ in $\sim_X^V$.

Intuitively, two states $w$ and $w'$ are equivalent up to set $X$, or $X$-equivalent, if and only if they satisfy the same atoms in $X$ (according to a given valuation $V$). The finiteness of $X$ is clearly not essential in the definition. It is assumed because, as we will see, each set $X$ will be taken to model a set of actions of some agent in a game form and sets of actions are always assumed to be finite.

We state the following simple fact without proof.

FACT 1. *(Properties of $\sim_P$) The following holds for any set of states $W$, valuation $V : \mathbf{P} \longrightarrow 2^W$ and finite sets $X, Y \subseteq \mathbf{P}$:*

(i) *$\sim_X$ is an equivalence relation on $W$;*

(ii) *if $X \subseteq Y$ then $\sim_Y \subseteq \sim_X$;*

(iii) *if $X$ is a singleton, $\sim_X$ induces a bipartition of $W$;*

(iv) *$\sim_X \cap \sim_Y = \sim_{X \cup Y}$;*

(v) *$\sim_\emptyset = W^2$.*

## 2.2 A modal logic of $\sim_X$

In this section we consider a simple modal language interpreted on relations $\sim_X$ and axiomatize its logic on the class of structures $(W, V)$. The key modal operator of the language will be $\langle X \rangle$, whose intuitive meaning is '$\varphi$ is the case in some state which is $X$-equivalent to the current one' or, to stress a ceteris paribus reading, '$\varphi$ is possible *all things expressed in $X$ being equal*'. We call the resulting logic *propositional equivalence ceteris paribus logic*, PECP in short.

### 2.2.1 Syntax of PECP.

Let $\mathbf{P}$ be a countable set of atomic propositions. The language $\mathcal{L}_{\mathsf{PECP}}(\mathbf{P})$ is defined by the following BNF:

$$\mathcal{L}_{\mathsf{PECP}}(\mathbf{P}) : \varphi ::= p \mid \neg\varphi \mid (\varphi \wedge \varphi) \mid \langle X \rangle \varphi$$

where $p$ ranges over $\mathbf{P}$ and $X$ is a finite subset of atomic propositions ($X \subseteq \mathbf{P}$ and $X$ finite). Note that as the set of finite subsets of atomic propositions is countable, the language $\mathcal{L}_{\mathsf{PECP}}(\mathbf{P})$ is also countable. The Boolean connectives $\top, \vee, \rightarrow, \leftrightarrow$ and the dual operators $[X]$ are defined as usual.

The set $SF(\varphi)$ of subformulas of a formula $\varphi$ is defined inductively as follows:

- $SF(p) = \{p\}$;
- $SF(\neg\varphi) = \{\neg\varphi\} \cup SF(\varphi)$;
- $SF(\varphi \wedge \psi) = \{\varphi \wedge \psi\} \cup SF(\varphi) \cup SF(\psi)$;
- $SF(\langle X \rangle \varphi) = \{\langle X \rangle \varphi\} \cup SF(\varphi)$.

We say that a signature $X$ appears in $\varphi$ if there exists a formula $\psi$ such that $\langle X \rangle \psi \in SF(\varphi)$.

### 2.2.2 Semantics of PECP

This is the class of models we will be working with:

DEFINITION 2. *(PECP-models) Given a countable set $\mathbf{P}$, a PECP-model for $\mathcal{L}_{\mathsf{PECP}}(\mathbf{P})$ is a tuple $\mathcal{M} = (W, V)$ where:*

- *$W$ is a non-empty set of states;*
- *$V : \mathbf{P} \longrightarrow 2^W$ is a valuation function.*

Intuitively, a PECP-model consists just of a state-space and a valuation function for a given set of atoms. The satisfaction relation is defined as follows:

DEFINITION 3. *(Satisfaction for PECP-models) Let $\mathcal{M} = (W, V)$ be an PECP-model for $\mathcal{L}_{\mathsf{PECP}}(\mathbf{P})$, $w \in W$ and $\varphi, \psi \in \mathcal{L}_{\mathsf{PECP}}(\mathbf{P})$:*

$$\mathcal{M}, w \models p \iff w \in V(p);$$
$$\mathcal{M}, w \models \neg\varphi \iff \mathcal{M}, w \not\models \varphi;$$
$$\mathcal{M}, w \models \varphi \wedge \psi \iff \mathcal{M}, w \models \varphi \text{ AND } \mathcal{M}, w \models \psi;$$
$$\mathcal{M}, w \models \langle X \rangle \varphi \iff \exists w' \in W : w \sim_X^V w' \text{ AND } \mathcal{M}, w' \models \varphi$$

*Formula $\varphi$ is PECP-satisfiable, if and only if there exists a model $\mathcal{M}$ and a state $w$ such that $\mathcal{M}, w \models \varphi$. Formula $\varphi$ is valid in $\mathcal{M}$, noted $\mathcal{M} \models \varphi$, if and only if for all $w \in W$, $\mathcal{M}, w \models \varphi$. Finally, $\varphi$ is PECP-valid, noted $\models_{\mathsf{PECP}} \varphi$, if and only if it is valid in all PECP-models. The logical consequence of formula $\varphi$ from a set of formulae, noted $\Phi \models_{\mathsf{PECP}} \varphi$, is defined as usual.*

So, modal operators are interpreted on the equivalence relations $\sim_X$ induced by the valuation of the model. It is worth observing that the logic of this class of models is not invariant under uniform substitution, suffice it to mention a validity such as $[\{p\}]p \vee [\{p\}]\neg p$.

### 2.2.3 Axiomatics of PECP

We can obtain an axiom system for PECP by a reduction technique. Let $X, Y$ range over finite elements of $2^\mathbf{P}$, $\varphi, \psi$ over $\mathcal{L}_{\mathsf{PECP}}(\mathbf{P})$, and $p$ over $\mathbf{P}$:

(P)  all tautologies of propositional calculus

(K)  $[\emptyset](\varphi \rightarrow \psi) \rightarrow ([\emptyset]\varphi \rightarrow [\emptyset]\psi)$

(T)  $\varphi \rightarrow \langle\emptyset\rangle\varphi$

(4)  $\langle\emptyset\rangle\langle\emptyset\rangle\varphi \rightarrow \langle\emptyset\rangle\varphi$

(5)  $\langle\emptyset\rangle\varphi \rightarrow [\emptyset]\langle\emptyset\rangle\varphi$

(Reduce)  $[X]\varphi \leftrightarrow \bigwedge_{\pi \subseteq X} \left( \left( \bigwedge_{p \in \pi} p \wedge \bigwedge_{p \in X \setminus \pi} \neg p \right) \rightarrow \right.$
$\left. [\emptyset]\left( \left( \bigwedge_{p \in \pi} p \wedge \bigwedge_{p \in X \setminus \pi} \neg p \right) \rightarrow \varphi \right) \right)$

And it is closed under the following inference rules ($\vdash_{\mathsf{PECP}}$ has its usual meaning):

(MP)  IF $\vdash_{\mathsf{PECP}} \varphi$ AND $\vdash_{\mathsf{PECP}} \varphi \rightarrow \psi$ THEN $\vdash_{\mathsf{PECP}} \psi$

(N)  IF $\vdash_{\mathsf{PECP}} \varphi$ THEN $\vdash_{\mathsf{PECP}} [\emptyset]\varphi$

The first thing to notice is that the system consists of S5 plus the Reduce axiom. Logic S5 is known to be sound and strongly complete for the class of models where the accessibility relation is the total relation $W^2$ [BdRV01], and modality $[\emptyset]$ is here axiomatized as one would axiomatize the global modality (cf. properties *(i)* and *(v)* in Fact 1).

Having said this, soundness and strong completeness of the above system are easy to establish. For soundness, it suffices to show that Reduce is PECP-valid, which follows straightforwardly from Definition 1. Intuitively, the axiom reduces $[X]\varphi$ by taking care of all the possible truth-value combinations of the atoms in $X$. If a given combination, e.g., $\left( \bigwedge_{p \in \pi} p \wedge \bigwedge_{p \in X \setminus \pi} \neg p \right)$, is true at a given state (for some $\pi$), then in all accessible states, if that combination is true, then $\varphi$ is also true.

To obtain completeness we proceed as customary in DEL [vKv07], by using axiom Reduce and the following rule of substitution of provable equivalents (REP) to remove the occurrences of those $\langle X \rangle$ and $[X]$ operators from formulae where $X \neq \emptyset$:

(REP)     IF $\vdash_{\mathsf{PECP}} \varphi \leftrightarrow \varphi'$ THEN $\vdash_{\mathsf{PECP}} \psi \leftrightarrow \psi[\varphi/\varphi']$

where $\psi[\varphi/\varphi']$ is the formula that results from $\psi$ by replacing zero or more occurrences of $\varphi$, in $\psi$, by $\varphi'$.

One can show that REP is derivable for every operator $[X]$ as follows: first one can show that each $[X]$ operator satisfies the Axiom K and the rule of necessitation N. Let us provide the syntactic proofs of this. For notational convenience we use the following abbreviation:

$$\widehat{\pi} \stackrel{\text{def}}{=} \left( \bigwedge_{p \in \pi} p \wedge \bigwedge_{p \in X \setminus \pi} \neg p \right)$$

Derivation of K for $[X]$:

1. $\vdash [X](\varphi \to \psi) \leftrightarrow \bigwedge_{\pi \subseteq X} (\widehat{\pi} \to [\emptyset](\widehat{\pi} \to (\varphi \to \psi)))$
   by Reduce

2. $\vdash (\widehat{\pi} \to (\varphi \to \psi)) \to ((\widehat{\pi} \to \varphi) \to (\widehat{\pi} \to \psi))$
   by P

3. $\vdash \bigwedge_{\pi \subseteq X} (\widehat{\pi} \to [\emptyset](\widehat{\pi} \to (\varphi \to \psi))) \to$
   $\bigwedge_{\pi \subseteq X} (\widehat{\pi} \to [\emptyset]((\widehat{\pi} \to \varphi) \to (\widehat{\pi} \to \psi)))$
   by P, 2 and RM for $[\emptyset]$ (if $\vdash \varphi \to \psi$ then $\vdash [\emptyset]\varphi \to [\emptyset]\psi$)

4. $\vdash \bigwedge_{\pi \subseteq X} (\widehat{\pi} \to [\emptyset]((\widehat{\pi} \to \varphi) \to (\widehat{\pi} \to \psi))) \to$
   $\bigwedge_{\pi \subseteq X} (\widehat{\pi} \to ([\emptyset](\widehat{\pi} \to \varphi) \to [\emptyset](\widehat{\pi} \to \psi)))$
   by K and P

5. $\vdash \bigwedge_{\pi \subseteq X} (\widehat{\pi} \to ([\emptyset](\widehat{\pi} \to \varphi) \to [\emptyset](\widehat{\pi} \to \psi))) \to$
   $(\bigwedge_{\pi \subseteq X} (\widehat{\pi} \to [\emptyset](\widehat{\pi} \to \varphi)) \to \bigwedge_{\pi \subseteq X} (\widehat{\pi} \to [\emptyset](\widehat{\pi} \to \psi)))$
   by P

6. $\vdash (\bigwedge_{\pi \subseteq X} (\widehat{\pi} \to [\emptyset](\widehat{\pi} \to \varphi)) \to$
   $\bigwedge_{\pi \subseteq X} (\widehat{\pi} \to [\emptyset](\widehat{\pi} \to \psi))) \leftrightarrow$
   $([X]\varphi \to [X]\psi)$
   by Reduce

7. $\vdash [X](\varphi \to \psi) \to ([X]\varphi \to [X]\psi)$
   from 1 and 3-6

Derivation of N for $[X]$:

1. $\vdash \varphi$
   hypothesis

2. $\vdash [\emptyset]\varphi$
   from 1 by N for $[\emptyset]$

3. $\vdash \bigwedge_{\pi \subseteq X} [\emptyset](\widehat{\pi} \to \varphi)$
   from 2 by the S5 theorem $[\emptyset]\varphi \to [\emptyset](\psi \to \varphi)$

4. $\vdash \bigwedge_{\pi \subseteq X} (\widehat{\pi} \to [\emptyset](\widehat{\pi} \to \varphi))$
   from 3 by P

5. $\vdash [X]\varphi$
   from 4 by Reduce and MP

Then one proves that REP is derivable by an induction routine analogous to the one used in [Che80, Th. 4.7].

We opted for this axiomatization in virtue of its simplicity, but alternative systems are of course possible. One in particular is worth mentioning. It first reduces $\langle p \rangle$ operators by axiom:

$$\langle p \rangle \varphi \leftrightarrow ((p \wedge \langle \emptyset \rangle (p \wedge \varphi)) \vee (\neg p \wedge \langle \emptyset \rangle (\neg p \wedge \varphi))) \quad (1)$$

This states that $\langle p \rangle \varphi$ is equivalent to either the case in which the current state satisfies $p$ and there exists a (possibly different) $p$-state where $\varphi$ is true, or the case where $\neg p$ is true and there exists a (possibly different) $\neg p$-state where $\varphi$ is true (recall property *(iii)* in Fact 1). Given the above reduction, one can then use axioms to enforce the appropriate behavior of $\sim_X$ relations where $X$ consists of more than one atom. To this aim, axioms can be used that are known to be canonical for properties *(ii)* and *(iv)* of Fact 1, namely:

$$\langle X \cup Y \rangle \varphi \to \langle X \rangle \varphi \quad (2)$$

$$\langle X \rangle i \wedge \langle Y \rangle i \to \langle X \cup Y \rangle i \quad (3)$$

where $i$ ranges over a set of nominals. A complete system could then be obtained by axiomatizing the behavior of nominals—through axioms and rules used in hybrid logic [AT06]. From that system, a named canonical model could be built (i.e., a canonical model where all maximal consistent sets contain exactly one nominal) where the axioms in Formulae 1-3 would enforce the desirable properties on the canonical relations.

## 2.3 Exponentially embedding PECP into S5

The property expressed by axiom Reduce enables a truth-preserving translation of PECP into S5. This translation is, however, such that the translated formula is exponentially larger by a tower of exponents of height equal to the modal depth of the original formula.

In this section we propose a translation that is single exponential and preserves satisfiability. Take the standard modal language $\mathcal{L}_\Box(\mathbf{P})$ with one modal operator $\Box$ defined on the set of atoms $\mathbf{P}$. S5-models are structures $\mathcal{M} = (W, V)$ where $W$ is a set of states, and $V : \mathbf{P} \longrightarrow 2^W$ a valuation function. Given an S5-model $\mathcal{M} = (W, V)$ and a state $w \in W$, the truth conditions are defined as follows:

$$\mathcal{M}, w \models \Box \varphi \iff \forall u \in W : \mathcal{M}, u \models \varphi$$

S5-satisfiability is defined as usual. It is possible to define an exponential truth-preserving reduction $tr : \mathcal{L}_{\mathsf{PECP}}(\mathbf{P}) \longrightarrow \mathcal{L}_\Box(\mathbf{P})$ as follows:

- $tr(\varphi_0) = p_{\varphi_0} \wedge \bigwedge_{\varphi \in SF(\varphi_0)} \Box (p_\varphi \leftrightarrow tr_1(\varphi))$

where $p_\varphi$ are fresh atomic propositions and $tr_1$ is defined as follows:

$$\begin{aligned}
tr_1(p) &= p \quad \text{FOR } p \in \mathbf{P} \\
tr_1(\neg\varphi) &= \neg tr_1(\varphi) \\
tr_1(\varphi \wedge \psi) &= tr_1(\varphi) \wedge tr_1(\psi) \\
tr_1([\emptyset]\varphi) &= \Box p_\varphi \\
tr_1([X]\varphi) &= \bigwedge_{\pi \subseteq X} \left( \left( \bigwedge_{p \in \pi} p \wedge \bigwedge_{p \in X \setminus \pi} \neg p \right) \rightarrow \right. \\
& \qquad \left. \Box \left( \left( \bigwedge_{p \in \pi} p \wedge \bigwedge_{p \in X \setminus \pi} \neg p \right) \rightarrow p_\varphi \right) \right)
\end{aligned}$$

Intuitively, the translation is designed to operate like axiom Reduce but avoiding exponential blow-up to pile up with the modal depth of the formula. The atomic propositions $p_\varphi$ in $tr_1([X]\varphi)$ avoid the non-elementary size of $tr(\varphi_0)$. The definition of $tr_1([\emptyset]\varphi)$ corresponds to the degenerated case of $tr_1([X]\varphi)$ where $X = \emptyset$. The following theorem states the satisfiability preservation. The proof is given in Appendix A.

THEOREM 1. *(tr preserves satisfiability) Let $\varphi_0$ be a PECP-formula. We have equivalence between $\varphi_0$ is PECP-satisfiable and $tr(\varphi_0)$ is S5-satisfiable.*

As a consequence, we also obtain the following result.

COROLLARY 1. *(Decidability) The satisfiability problem for PECP is decidable and in NEXPTIME.*

PROOF. The satisfiability problem for S5 is decidable and in NP [BdRV01]. The result follows from Theorem 1 and a decision procedure may work as follows: in order to check that $\varphi$ is satisfiable we compute the formula $tr(\varphi)$ and we apply a NP-decision procedure to check whether $tr(\varphi)$ is S5-satisfiable or not. □

Notice that if the cardinality of each $X$ that appears in operators $[X]$ of $\varphi$ is bounded by a fixed integer, then the translation $tr$ becomes polynomial in the size of $\varphi$. Thus, as S5-satisfiability problem is NP-complete, the PECP-satisfiability problem with a bounded cardinality restrictions over set of atomic propositions in modal operators is in NP. As it is trivially NP-hard, it is NP-complete.

In Section 3, we will embed the atemporal version of STIT (the logic of *seeing to it that*) into PECP thereby obtaining lower bounds results.

## 2.4 PECP and modal ceteris paribus logics

Before moving to the next section, we briefly compare PECP with two works in the modal logic of ceteris paribus reasoning: release logic, and the logic of ceteris paribus preference.

Release logic has been introduced and studied in [KM03, KM00] in order to provide a modal logic characterization of a general notion of irrelevancy. Modal operators in release logic are S5 operators indexed by subsets of a finite set Iss of abstract elements denoting the issues that are taken to be irrelevant, or that can be *released*, while evaluating the formula in the scope of the operator. A release model is therefore a tuple $(W, \{\sim^r_X\}_{X \subseteq \texttt{Iss}}, V)$ where all $\sim^r_X$ are equivalence relations with the additional constraint that if $X \subseteq Y$ then $\sim^r_X \subseteq \sim^r_Y$, that is, by releasing more issues one obtains a more granular relation. This is, more precisely, the semantics of release operators:

$$\mathcal{M}, w \models \Diamond_X \varphi \iff \exists w' \in W : w \sim^r_X w' \text{ AND } \mathcal{M}, w' \models \varphi$$

where $X \subseteq \texttt{Iss}$.

One can easily observe that, by Fact 1 (clause *(ii)*), PECP models are release models where $\texttt{Iss} = \mathbf{P}$ and where the release relation $\sim^r_X = \sim_{-X}$. Vice versa, for $\texttt{Iss} = \mathbf{P}$, not all release models are PECP models. As a consequence, the logic of $\langle -X \rangle$ operators in PECP is a conservative extension of the logic of $\Diamond_X$ release operators.

Preference logic has also long been concerned with so-called ceteris paribus preferences, that is, preferences incorporating an "all other things being equal" condition. A first logical analysis of such preferences dates back to [Von63], where dyadic modal operators are studied representing statements like '$\varphi$ is preferred to $\psi$, ceteris paribus'. More recently, [vBGR09] has provided a modal logic of ceteris paribus preferences based on standard unary modal operators. Leaving the preferential component of such logic aside, its ceteris paribus fragment concerns sentences of the form $\langle \Gamma \rangle \varphi$ whose intuitive meaning is 'there exists a state which is equivalent to the evaluation state with respect to all the formulae in the finite set $\Gamma$ and which satisfies $\varphi$', where the formulae in $\Gamma$ are drawn from the full language. It is easy to see that logic PECP is, in fact, the fragment of the ceteris paribus logic where $\Gamma$ is allowed to consist only of a finite set of atoms.

## 3. PECP EMBEDDING OF ATEMPORAL STIT

In this section, we investigate the possibility of embedding the logic of agency STIT into PECP. STIT logic (the logic of *seeing to it that*) [BPX01, Hor01] is one of the most prominent logical accounts of agency. It is the logic of constructions of the form "agent $i$ (or group $J$) sees to it that $\varphi$". STIT has a non-standard modal semantics based on the concepts of *moment* and *history*. However, as shown by [BHT08, HS08], the basic STIT language without temporal operators can be 'simulated' in a standard Kripke semantics.

### 3.1 Atemporal group STIT

First let us recall the syntax and the semantics of atemporal group STIT. The language of this logic is built from a countable set of atomic propositions $\mathbf{P}$ and a finite set of agents $AGT = \{1, \ldots, n\}$ and is defined by the following BNF:

$$\mathcal{L}_{G-\mathsf{STIT}}(\mathbf{P}, AGT) : \varphi ::= p \mid \neg\varphi \mid (\varphi \wedge \varphi) \mid [J : stit]\varphi$$

where $p$ ranges over $\mathbf{P}$ and $J$ ranges over $2^{AGT}$. The construction $[J : stit]\varphi$ is read "group $J$ sees to it that $\varphi$ is true regardless of what the other agents choose". We define the dual operator $\langle J : stit \rangle \varphi \stackrel{\text{def}}{=} \neg[J : stit]\neg\varphi$. When $J = \emptyset$, the construction $[\emptyset : stit]\varphi$ is read "$\varphi$ is true regardless of what every agent chooses" or simply "$\varphi$ is necessarily true".

DEFINITION 4 (STIT-KRIPKE MODEL [HS08]). *A STIT-Kripke model $\mathcal{M} = (W, \{R_J\}_{J \subseteq AGT}, V)$ is a 3-tuple where:*

- *$W$ is a non-empty set of worlds;*
- *for all $J \subseteq AGT$, $R_J$ is an equivalence relation such that:*

i) $R_J \subseteq R_\emptyset$;

ii) $R_J = \bigcap_{j \in J} R_{\{j\}}$;

iii) for all $w, u_1, \ldots, u_n \in W$, if $u_1 \in R_\emptyset(w), \ldots, u_n \in R_\emptyset(w)$ then $\bigcap_{1 \leq j \leq n} R_{\{j\}}(u_j) \neq \emptyset$;

- $V : \mathbf{P} \to 2^W$ is a valuation function for atomic propositions;

with $R_J(w) = \{u \in W : (w,u) \in R_J\}$ for any $J \in 2^{AGT}$.

The partition induced by the equivalence relation $R_J$ is the set of possible choices of the group $J$.[1] Indeed, in STIT a choice of a group $J$ at a given world $w$ is identified with the set of possible worlds $R_J(w)$. We call $R_J(w)$ the set of possible outcomes of group $J$'s choice at world $w$, in the sense that group $J$'s current choice at $w$ forces the possible worlds to be in $R_J(w)$. The set $R_\emptyset(w)$ is simply the set of possible outcomes at $w$, or said differently, the set of outcomes of the current game at $w$. According to Condition *(i)*, the set of possible outcomes of a group $J$'s choice is a subset of the set of possible outcomes. Condition *(ii)*, called *additivity*, means that the choices of the agents in a group $J$ is made up of the choices of each individual agent and no more. Condition *(iii)* corresponds to the property of *independence of agents*: whatever each agent decides to do, the set of outcomes corresponding to the joint action of all agents is non-empty. More intuitively, this means that agents can never be deprived of choices due to the choices made by other agents. In [LS11] we supposed determinism for the group $AGT$, that is to say that the set of outcomes corresponding to a joint action of all agents is a singleton. Horty's group STIT logic [Hor01] does not suppose this. Here we deal with Horty's version of STIT. So a STIT model is a game form in which a joint action of all agents might determine more than one outcome.

EXAMPLE 1. *The tuple $\mathcal{M} = (W, R_\emptyset, R_{\{1\}}, R_{\{2\}}, R_{\{1,2\}}, V)$ defined by:*

- $W = \{w, u, v, r, s, t, z\}$;

- $R_\emptyset = W \times W$;

- $R_{\{1\}} = \{w, u, v\}^2 \cup \{r, s\}^2 \cup \{t, z\}^2$;

- $R_{\{2\}} = \{w, r, t\}^2 \cup \{u, v, s, z\}^2$;

- $R_{\{1,2\}} = \{(w,w), (r,r), (s,s), (t,t), (z,z),$
  $(u,u), (v,v), (u,v), (v,u)\}$;

- *for all $p \in \mathbf{P}$, $V(p) = \emptyset$.*

*is a STIT-Kripke model. Figure 1 shows the model $\mathcal{M}$. The equivalence classes induced by the equivalence relation $R_{\{1\}}$ are represented by ellipses and correspond to the choices of agent 1. The equivalence classes induced by the equivalence relation $R_{\{2\}}$ are represented by rectangles and correspond to the choices of agent 2. The choice of group $\{1,2\}$ at a given world is determined by the intersection of the choice of agent 1 and the choice of agent 2 at this world. For example, the choice of agent 1 at world $u$ is $\{w, u, v\}$ whereas the choice*

---

[1] One can also see the partition induced by the equivalence relation $R_j$ as the set of actions that agent $j$ can *try*, where the notion of *trying* corresponds to the notion of *volition* studied in philosophy of action [O'S74, McC74].

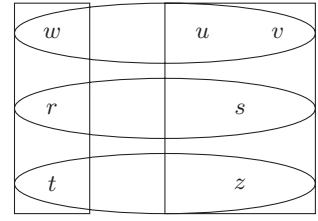

**Figure 1: The model $\mathcal{M}$**

*of agent 2 at world $u$ is $\{u, v, s, z\}$. The choice of group $\{1,2\}$ at $u$ is $\{u, v\}$. Note that Condition* (iii) *of Definition 4 ensures that for any choice of agent 1 and for any choice of agent 2 the intersection between these two choices is non-empty. That is, for any equivalence class induced by the relation $R_{\{1\}}$ and for any equivalence class induced by the relation $R_{\{2\}}$, the intersection between these two equivalence classes is non-empty.*

Given a STIT-Kripke model $\mathcal{M} = (W, \{R_J\}_{J \subseteq AGT}, V)$ and a world $w$ in $\mathcal{M}$, the truth conditions of STIT formulae are the following:

$$\begin{aligned}
\mathcal{M}, w \models p &\iff w \in V(p); \\
\mathcal{M}, w \models \neg\varphi &\iff \mathcal{M}, w \not\models \varphi; \\
\mathcal{M}, w \models \varphi \land \psi &\iff \mathcal{M}, w \models \varphi \text{ AND } \mathcal{M}, w \models \psi; \\
\mathcal{M}, w \models [J : stit]\varphi &\iff \forall v \in R_J(w) : \mathcal{M}, v \models \varphi
\end{aligned}$$

where $R_J(w) = \{u \in W \mid (w,u) \in R_J\}$.

We are not able to embed group STIT into PECP because of many reasons. The first one is that the group STIT satisfiability problem is undecidable if there are more than 3 agents [HS08].[2] The second one is that group STIT does not have the finite model property. Indeed in [HS08], a translation from the product logic $S5^n$ to group STIT logic is given and as $S5^n$ does not have the finite model property [GKWZ03], atemporal group STIT will also not have it. On the contrary PECP inherits the finite model property from S5. Indeed, if a formula $\varphi$ is PECP-satisfiable, Theorem 1 says that $tr(\varphi)$ is S5-satisfiable. But as S5 has the polynomial model property, there exists a polynomial-sized S5-model for $tr(\varphi)$ in the size of $tr(\varphi)$. In other words, there exists an exponential S5-model for $tr(\varphi)$ in the size of $\varphi$. Theorem 1 ensures that there exists an exponential PECP-model for $\varphi$ in the size of $\varphi$.

We will nevertheless embed a variant of group STIT under the assumption that every agent has a finite and bounded number of actions in his repertoire. For every agent $j$, a $R_j$-equivalence class $R_j(u)$ corresponds to an action of agent $j$. We say that agent $j$ has $k_j$ actions in a STIT model if and only if there are exactly $k_j$ $R_j$-equivalence classes in $\mathcal{M}$.

The game structure in STIT-models should be enforced in PECP-models. That is why we introduce special atomic propositions to encode the game structure. Without loss of generality, we assume that the set $\mathbf{P}$ contains special atomic propositions $\mathfrak{rep}_1^j, \mathfrak{rep}_2^j, \ldots$ for all agents $j$ which are used to represent the actions of the agents. Let $k$ be the maximal number of actions: $k = max_{j \in AGT} k_j$. For every agent, we represent its actions by numbers $\ell$ in $\{0, \ldots, k-1\}$ and some atomic propositions encode the binary representation of $\ell$. Let $m$ be an integer that represents the number of digits we need to represent an action. For instance let $m = \lceil log_2 k \rceil$ (the ceiling of the logarithm of $k$). For a given agent $j$, $\mathfrak{R}_m^j = \{\mathfrak{rep}_1^j, \ldots, \mathfrak{rep}_m^j\}$ is the set atomic propositions that represent the binary digits of an action of agent $j$. We suppose that if $j \neq i$ then $\mathfrak{R}_m^j \cap \mathfrak{R}_m^i = \emptyset$.

---

[2] See [LS11] for a study of some decidable fragments of group STIT.

EXAMPLE 2. *For example, in the model of Example 1, agent 1 has $k_1 = 3$ actions and agent 2 has $k_2 = 2$ actions. So $k = 3$ and $m = \lceil log_2 3 \rceil = 2$. We have $\mathfrak{R}_m^1 = \{\mathfrak{rep}_1^1, \mathfrak{rep}_2^1\}$ and $\mathfrak{R}_m^2 = \{\mathfrak{rep}_1^2, \mathfrak{rep}_2^2\}$. Then for instance, we may represent the action of agent 1 corresponding to $R_{\{1\}}(w) = \{w, u, v\}$ by the valuation $\neg \mathfrak{rep}_1^1 \wedge \neg \mathfrak{rep}_2^1$, the action of agent 1 corresponding to $\{r, s\}$ by $\mathfrak{rep}_1^1 \wedge \neg \mathfrak{rep}_2^1$, the action of agent 1 corresponding to $\{t, z\}$ by $\neg \mathfrak{rep}_1^1 \wedge \mathfrak{rep}_2^1$, the action of agent 2 corresponding to $\{w, r, t\}$ by $\neg \mathfrak{rep}_1^2 \wedge \neg \mathfrak{rep}_2^2$ and the action of agent 2 corresponding to $\{u, v, s, z\}$ by $\mathfrak{rep}_1^2 \wedge \neg \mathfrak{rep}_2^2$.*

Let $\mathfrak{R}_m = \bigcup_{j \in AGT} \mathfrak{R}_m^j$ be the set of all atomic propositions used to denote actions. Let us define the following PECP formula:

$$GRID_m \stackrel{\text{def}}{=} \bigwedge_{x \in \mathfrak{R}_m} [\emptyset]((x \to \langle \mathfrak{R}_m \setminus \{x\} \rangle \neg x) \wedge$$
$$(\neg x \to \langle \mathfrak{R}_m \setminus \{x\} \rangle x))$$

This formula enforces a STIT model to contain all possible valuations over $\mathfrak{R}_m$. A model that satisfies $GRID_m$ is then interpreted as a game form where each valuation of $\mathfrak{R}_m^j$ represents an action of player $j$.

We now define a translation from $\mathcal{L}_{G-\text{STIT}}$ to $\mathcal{L}_{\text{PECP}}(\mathbf{P})$ as follows:

$$\begin{aligned}
tr_2(p) &= p \quad \text{FOR } p \in \mathbf{P} \\
tr_2(\neg \varphi) &= \neg tr_2(\varphi) \\
tr_2(\varphi \wedge \psi) &= tr_2(\varphi) \wedge tr_2(\psi) \\
tr_2([J : stit]\varphi) &= [\bigcup_{j \in J} \mathfrak{R}_m^j] tr_2(\varphi)
\end{aligned}$$

The translation $tr_2$ should be parameterized by $m$. For notational convenience, in what follows we write $tr_2$ instead of $tr_2^m$ leaving implicit the parameter $m$.

The set $\bigcup_{j \in J} \mathfrak{R}_m^j$ represents all the atomic propositions used to represented actions of the coalition $J$. We then have the following theorem whose proof is given in Appendix B at the end of the paper.

THEOREM 2. *Let us consider a group STIT formula $\varphi$. Let $m$ be an integer. Then the following items are equivalent:*

1. *$\varphi$ is STIT-satisfiable in a STIT-model where each agent has at most $2^m$ actions;*

2. *$\varphi$ is STIT-satisfiable in a STIT-model where each agent has exactly $2^m$ actions;*

3. *$GRID_m \wedge tr_2(\varphi)$ is PECP-satisfiable.*

### 3.2 Atemporal individual STIT

In this subsection, we consider the following fragment of STIT called atemporal individual STIT [3]:

$$\mathcal{L}_{I-\text{STIT}}(\mathbf{P}, AGT) : \varphi ::= p \mid \neg \varphi \mid (\varphi \wedge \varphi) \mid [\{j\} : stit]\varphi$$

where $p$ ranges over $\mathbf{P}$ and $j$ ranges over $AGT$.

This fragment of STIT, axiomatized by Xu in [Xu98], has the exponential finite model property (see Lemma 7 in [BHT08]). Moreover, as the following theorem highlights, it can be embedded in the logic PECP.

THEOREM 3. *Let us consider a STIT formula $\varphi$ of the individual STIT fragment. Let $m$ be the length of $\varphi$. Then the following three items are equivalent:*

1. *$\varphi$ is STIT-satisfiable*

2. *$\varphi$ is STIT-satisfiable in a model where each agent has at most $2^m$ actions;*

3. *$GRID_m \wedge tr_2(\varphi)$ is PECP-satisfiable.*

PROOF. $\boxed{1 \Rightarrow 2}$ Consider a STIT formula $\varphi$ of the individual STIT fragment. If $\varphi$ is STIT-satisfiable and $m$ is the length of $\varphi$, then $\varphi$ is STIT-satisfiable in a model where there are at most $2^m$ worlds (see Lemma 7 in [BHT08]). This implies that there are at most $2^m$ actions in that model.

The implications $2 \Rightarrow 3$ and $3 \Rightarrow 1$ come from Theorem 2. □

Thanks to Theorem 3, we reduce the NEXPTIME-complete satisfiability problem of individual STIT [BHT08] to the PECP-satisfiability problem. As the reduction is polynomial, we obtain the following lower bound complexity result for the PECP-satisfiability problem.

COROLLARY 2. *The PECP-satisfiability problem is NEXPTIME-hard.*

### 3.3 Group STIT where coalitions are nested

In this subsection we address the satisfiability problem of the fragment of PECP consisting of formulae $\varphi$ of $\mathcal{L}_{\text{PECP}}$ such that the sets of atomic propositions that appear in any operator $[X]$ occurring in $\varphi$ form a linear set of sets of atomic propositions. More formally, if $[X]$ and $[X']$ are two operators occurring in $\varphi$ then either $X \subseteq X'$ or $X' \subseteq X$. For instance, the formula $[\{p, q\}](\psi \wedge [\{p\}][\{p, q, r, s\}]\varphi)$ belongs to the fragment because $\{p\} \subseteq \{p, q\} \subseteq \{p, q, r, s\}$. On the contrary, the formula $[\{p\}]p \wedge [\{q\}]p$ is not an element of this fragment of PECP.

We call the satisfiability problem of this fragment of PECP the PECP-nested satisfiability problem. Due to the embedding proposed in Theorem 2 of STIT into PECP, we provide the following lower bound complexity result for the PECP-nested satisfiability problem. The proof is given in Appendix C.

THEOREM 4. *The PECP-nested satisfiability problem is PSPACE-hard.*

The following theorem provides an upper bound complexity result for this fragment of PECP. The proof is given in Appendix D.

---

[3] Some authors ([Bro08a, Wan06]) use the term 'multi-agent STIT' to designate the logic where operators are of the form $[\{j\} : stit]$. Here we prefer to use the more explicit term 'individual STIT' as in [HS08].

THEOREM 5. *The* PECP-*nested satisfiability problem is in* PSPACE.

This concludes our analysis of STIT logics via PECP. In the next section we move to the coalition logic of propositional control.

## 4. RELATING STIT WITH CL−PC, AND PECP WITH CL−PC

In this section we study the relationships between PECP, atemporal 'bounded' group STIT, and another well-known game logic, the logic CL−PC (*coalition logic of propositional control*).[4] Specifically, we show that CL−PC can be embedded into atemporal 'bounded' group STIT and, by the fact that atemporal 'bounded' group STIT can be embedded into PECP (Section 3.1), we indirectly show that CL−PC can be embedded into PECP.

CL−PC was introduced by [vdHW05] as a formal language for reasoning about capabilities of agents and coalitions in multiagent environments. In this logic the notion of capability is modeled by means of the concept of *control*. In particular, it is assumed that each agent $i$ is associated with a specific finite subset $\mathbf{P}_i$ of the finite set of all propositions $\mathbf{P}$. $\mathbf{P}_i$ is the set of propositions *controlled* by the agent $i$. That is, the agent $i$ has the ability to assign a (truth) value to each proposition in $\mathbf{P}_i$ but cannot affect the truth values of the propositions in $\mathbf{P} \setminus \mathbf{P}_i$. In the variant of CL−PC studied by [vdHW05] it is also assumed that control over propositions is exclusive, that is, two agents cannot control the same proposition (i.e., if $i \neq j$ then $\mathbf{P}_i \cap \mathbf{P}_j = \emptyset$). Moreover, it is assumed that control over propositions is complete, that is, every proposition is controlled by at least one agent (i.e., for every $p \in \mathbf{P}$ there exists an agent $i$ such that $p \in \mathbf{P}_i$).

The preceding concepts and assumptions are precisely formulated in the following section, which illustrates the syntax and the formal semantics of CL−PC.

### 4.1 Syntax and semantics of CL−PC

The *language of* CL−PC is built from a *finite* set of atomic propositions $\mathbf{P}$ and a finite set of agents $AGT = \{1, \ldots, n\}$, and is defined by the following BNF:

$$\mathcal{L}_{\mathsf{CL-PC}}(\mathbf{P}, AGT) : \varphi ::= p \mid \neg\varphi \mid (\varphi \wedge \varphi) \mid \Diamond_J \varphi$$

where $p$ ranges over $\mathbf{P}$ and $J$ ranges over $2^{AGT}$. Operator $\Diamond_J$ is called *cooperation modality*, and the construction $\Diamond_J \varphi$ means that "group $J$ has the contigent ability to achieve $\varphi$".

DEFINITION 5 (CL−PC MODEL). *A model for* CL−PC *is a tuple* $\mathcal{M} = (\mathbf{P}_1, \ldots, \mathbf{P}_n, X)$ *where:*

- $\mathbf{P}_1, \ldots, \mathbf{P}_n$ *is a partition of* $\mathbf{P}$ *among the agents in* $AGT$;

- $X \subseteq \mathbf{P}$ *is the set of propositions which are true in the initial state.*

For every group of agents $J \subseteq AGT$, let $\mathbf{P}_J = \bigcup_{i \in J} \mathbf{P}_i$ be the set of atomic propositions controlled by the group $J$. Moreover, for every group $J \subseteq AGT$ and for every set of atomic propositions $X \subseteq \mathbf{P}$, let $X_J = X \cap \mathbf{P}_J$ be the set of atomic propositions in $X$ controlled by the group $J$. Sets $X_J$ are called $J$-valuations.

Given a CL−PC model $\mathcal{M} = (\mathbf{P}_1, \ldots, \mathbf{P}_n, X)$, the truth conditions of CL−PC formulae are the following:

$$\begin{aligned}
\mathcal{M} \models p &\iff p \in X; \\
\mathcal{M} \models \neg\varphi &\iff \mathcal{M} \not\models \varphi; \\
\mathcal{M} \models \varphi \wedge \psi &\iff \mathcal{M} \models \varphi \text{ AND } \mathcal{M} \models \psi; \\
\mathcal{M} \models \Diamond_J \varphi &\iff \exists X'_J \subseteq \mathbf{P}_J : \mathcal{M} \bigoplus X'_J \models \varphi
\end{aligned}$$

where $\mathcal{M} \bigoplus X'_J$ is the CL−PC model $(\mathbf{P}_1, \ldots, \mathbf{P}_n, X'')$ such that:

$$\begin{aligned}
X''_{AGT \setminus J} &= X_{AGT \setminus J} \\
X''_J &= X'_J
\end{aligned}$$

That is, $\Diamond_J \varphi$ is true at a given model $\mathcal{M}$ if and only if, the coalition $J$ can change the truth values of the atoms that it controls in such a way that $\varphi$ will be true afterwards (i.e., given the actual truth-value combination of the atoms which are not controlled by $J$, there exists a truth-value combination of the atoms controlled by $J$ which ensures $\varphi$).

Let us illustrate the CL−PC semantics with an example.

EXAMPLE 3. *Let* $AGT = \{1, 2, 3\}$, $\mathbf{P} = \{p, q, r\}$, $\mathbf{P}_1 = \{p\}$, $\mathbf{P}_2 = \{q\}$ *and* $\mathbf{P}_3 = \{r\}$.

*Consider the* CL−PC *model* $\mathcal{M} = (\mathbf{P}_1, \mathbf{P}_2,, \mathbf{P}_3, \{r\})$. *We have that:*

$$\mathcal{M} \models \Diamond_{\{1,2\}}((p \wedge q \wedge r) \vee (p \wedge \neg q \wedge r)).$$

*Indeed, there exists a set of atoms* $X'_{\{1,2\}} \subseteq \mathbf{P}_{\{1,2\}}$ *controlled by* $\{1, 2\}$ *such that* $\mathcal{M} \bigoplus X'_{\{1,2\}} \models ((p \wedge q \wedge r) \vee (p \wedge \neg q \wedge r))$.

*For example, we have* $\{p\} \subseteq \mathbf{P}_{\{1,2\}}$ *and* $(\mathbf{P}_1, \mathbf{P}_2, \mathbf{P}_3, \{p, r\}) \models ((p \wedge q \wedge r) \vee (p \wedge \neg q \wedge r))$, *where* $(\mathbf{P}_1, \mathbf{P}_2, \mathbf{P}_3, \{p, r\}) = \mathcal{M} \bigoplus \{p\}$.

### 4.2 Embedding CL−PC into STIT

The aim of this section is to provide an embedding of CL−PC into the variant of atemporal group STIT with bounded choices (atemporal 'bounded' group STIT) that have been presented in Section 3.1.

Let us provide the following STIT formulae which catpure four basic assumptions of CL−PC:

$$EXC^+ \stackrel{\text{def}}{=} \bigwedge_{p \in \mathbf{P}} \bigwedge_{i,j \in AGT : i \neq j} (\langle \emptyset : stit \rangle [\{i\} : stit]p \to \neg \langle \emptyset : stit \rangle [\{j\} : stit]p)$$

$$EXC^- \stackrel{\text{def}}{=} \bigwedge_{p \in \mathbf{P}} \bigwedge_{i,j \in AGT : i \neq j} (\langle \emptyset : stit \rangle [\{i\} : stit]p \to \neg \langle \emptyset : stit \rangle [\{j\} : stit] \neg p)$$

$$COMPL \stackrel{\text{def}}{=} \bigwedge_{p \in \mathbf{P}} \bigvee_{i \in AGT} [\emptyset : stit]([\{i\} : stit]p \vee [\{i\} : stit] \neg p)$$

$$GRID^* \stackrel{\text{def}}{=} \bigwedge_{X \subseteq \mathbf{P}} \langle \emptyset : stit \rangle (\bigwedge_{p \in X} p \wedge \bigwedge_{p \in \mathbf{P} \setminus X} \neg p)$$

Formulae $EXC^+$ and $EXC^-$ mean that control over atomic propositions in $\mathbf{P}$ is exclusive (i.e., there is no proposition in **P** which can be forced to be true or false by more than one agent), whereas formula $COMPL$ means that exercise of control over atomic propositions in **P** is complete (i.e., for every proposition in **P** there exists at least one agent who either forces it to be true or forces it to be false). Finally, formula $GRID^*$ means that all the possible truth-value combinations of the atomic propositions in **P** are possible. Note that $EXC^+$, $EXC^-$, $COMPL$ and $GRID^*$ are well-formed STIT formulae because of the assumption that the set **P** is finite.[5]

---

[4] In [Ger06] generalizations of some of the assumptions underlying CL−PC have been studied. Here we only consider the original version of CL−PC proposed by van der Hoek & Wooldridge.

We define the following translation from $\mathcal{L}_{\mathsf{CL-PC}}(\mathbf{P}, AGT)$ to $\mathcal{L}_{\mathsf{STIT}}(\mathbf{P}, AGT)$:

$$\begin{aligned} tr_3(p) &= p \text{ for } p \in \mathbf{P} \\ tr_3(\neg \varphi) &= \neg tr_3(\varphi) \\ tr_3(\varphi \wedge \psi) &= tr_3(\varphi) \wedge tr_3(\psi) \\ tr_3(\Diamond_J \varphi) &= \langle AGT \setminus J : stit \rangle tr_3(\varphi) \end{aligned}$$

The following theorem highlights that 'bounded' group STIT embeds CL−PC. The proof is given in Appendix E.

THEOREM 6. *Let $m = |\mathbf{P}|$. Then, a CL−PC formula $\varphi$ is CL−PC-satisfiable if and only if $(EXC^+ \wedge EXC^- \wedge COMPL \wedge GRID^*) \wedge tr_3(\varphi)$ is satisfiable in a STIT model where each agent has at most $2^m$ actions.*

As PECP embeds atemporal 'bounded' group STIT (Theorem 2 in Section 3.1), from Theorem 6 it follows that PECP also embeds CL−PC. Indeed, given a CL−PC-satisfiable formula $\varphi$, one can use the translation $tr_2$ given in Section 3.1 in order to find a corresponding STIT formula which is STIT-satisfiable. Then, one uses the preceding translation $tr_3$ in order to find a corresponding PECP formula which is PECP-satisfiable.

COROLLARY 3. *Let $m = |\mathbf{P}|$. Then, a CL−PC formula $\varphi$ is CL−PC-satisfiable if and only if $GRID_m \wedge tr_2((EXC^+ \wedge EXC^- \wedge COMPL \wedge GRID^*) \wedge tr_3(\varphi))$ is PECP-satisfiable.*

## 5. CONCLUSIONS

The paper has introduced a modal logic that arises by interpreting modal operators on the equivalence relations induced by finite sets of propositional atoms. This logic, called PECP, has been axiomatized, embedded (exponentially) into S5, and its relation to existing formalisms has been briefly discussed. PECP has then been used as a tool to compare two logics of agency and games—atemporal STIT and the coalitional logic of propositional control CL−PC—showing that CL−PC can be embedded in STIT and that, in turn, STIT can be embedded in PECP. These embedding preserve satisfiability and the paper has taken stock of them to provide a complexity analysis of logic PECP.

Moreover, via logic S5, one can easily show that embeddings in the other directions are also possible. S5, we have seen, embeds PECP, but is also directly embeddable in all the mentioned logic, which all contain the universal modality, in the following forms: $\langle \emptyset \rangle$ in PECP, $\langle AGT \setminus \emptyset : stit \rangle$ in atemporal STIT and $\Diamond_{AGT}$ in CL−PC. All in all, this illustrates a nice uniformity in the logical tools that seem to be needed to talk about $\alpha$-effectivity and, we believe, that

PECP offers a good paradigm for systematizing existing logics of game forms.

Directions of future work are manifold. First of all, we plan to look for principled generalizations of some of the assumptions underlying the logics studied: e.g., independence of agents in STIT such as "agent $j$ and agent $i$ are independent as far as the set of atomic propositions $X$ is concerned", restriction to control over atomic propositions in CL−PC. Secondly, we intend to push further our study of the relationship between ceteris paribus logics and existing logics of agency and cooperation including the logic of "bringing it about that" [GR05], the logic STIT with time [Bro08b, Lor12] and Coalition Logic [Pau02]. Finally, in this paper we have shown that PECP and atemporal individual STIT have the same high complexity of the satisfiability problem when we consider the whole languages. The study of efficient syntactic fragments is then important and we intend to pursue this study in parallel both for PECP and for atemporal individual STIT. We expect that several complexity results about fragments of atemporal STIT may be transferred to fragments of PECP and viceversa.

## 6. REFERENCES

[AT06] C. Areces and B. Ten Cate. Hybrid logics. In P. Blackburn, J. van Benthem, and F. Wolter, editors, *Handbook of Modal Logic*, pages 821–868. Elsevier, 2006.

[BdRV01] P. Blackburn, M. de Rijke, and Y. Venema. *Modal Logic*. Cambridge University Press, Cambridge, 2001.

[BHT08] P. Balbiani, A. Herzig, and N. Troquard. Alternative axiomatics and complexity of deliberative stit theories. *Journal of Philosophical Logic*, 37(4):387–406, 2008.

[BPX01] N.D. Belnap, M. Perloff, and M. Xu. *Facing the future: agents and choices in our indeterminist world*. Oxford University Press, USA, 2001.

[Bro08a] J. Broersen. A complete STIT logic for knowledge and action, and some of its applications. In *Proceedings of the 6th International Workshop on Declarative Agent Languages and Technologies (DALT 2008)*, volume 5397 of *LNCS*, pages 47–59. Springer-Verlag, 2008.

[Bro08b] J. Broersen. A logical analysis of the interaction between 'obligation-to-do' and 'knowingly doing'. In *Proceedings of the Ninth International Conference on Deontic Logic in Computer Science (DEON'08)*, volume 5076 of *LNCS*, pages 140–154. Springer-Verlag, 2008.

[Che80] B. F. Chellas. *Modal Logic. An Introduction.* Cambridge University Press, Cambridge, 1980.

[Ger06] J. Gerbrandy. Logics of propositional control. In *Proceedings of AAMAS'06*, pages 193–200. ACM, 2006.

[GKWZ03] D. M. Gabbay, A. Kurucz, F. Wolter, and M. Zakharyaschev. *Many-dimensional modal logics: theory and applications*. Elsevier, 2003.

[GR05] G. Governatori and A. Rotolo. On the axiomatization of Elgesem's logic of agency and ability. *Journal of Philosophical Logic*, 34:403–431, 2005.

---

[5]This assumption is also made by van der Hoek & Wooldridge in [vdHW05].

[Hor01]  J. F. Horty. *Agency and Deontic Logic*. Oxford University Press, Oxford, 2001.

[HS08]  A. Herzig and F. Schwarzentruber. Properties of logics of individual and group agency. *Advances in modal logic*, 7:133–149, 2008.

[KM00]  J. Krabbendam and J.-J.Ch. Meyer. Release logics for temporalizing dynamic logic, orthogonalising modal logics. In M. Barringer, M. Fisher, D. Gabbay, and G. Gough, editors, *Advances in Temporal Logic*, pages 21–45. Kluwer Academic Publisher, 2000.

[KM03]  J. Krabbendam and J.-J. Ch. Meyer. Contextual deontic logics. In P. McNamara and H. Prakken, editors, *Norms, Logics and Information Systems*, pages 347–362, Amsterdam, 2003. IOS Press.

[Lor12]  E. Lorini. A STIT-logic analysis of commitment and its dynamics. *Journal of Applied Logic*, 2012. to appear.

[LS11]  E. Lorini and F. Schwarzentruber. A logic for reasoning about counterfactual emotions. *Artificial Intelligence*, 175(3-4):814–847, 2011.

[McC74]  H. J. McCann. Volition and basic action. *The Philosophical Review*, 83:451–473, 1974.

[MP82]  H. Moulin and B. Peleg. Cores of effectivity functions and implementation theory. *Journal of Mathematical Economics*, 10:115–145, 1982.

[O'S74]  B. O'Shaughnessy. Trying (as the mental pineal gland). *The Journal of Philosophy*, 70:365–386, 1974.

[Pau02]  M. Pauly. A modal logic for coalitional power in games. *Journal of Logic and Computation*, 12(1):149–166, 2002.

[Sch12]  F. Schwarzentruber. Complexity results of stit fragments. *Studia logica*, 2012. to appear.

[vBGR09]  J. van Benthem, P. Girard, and O. Roy. Everything else being equal: A modal logic for ceteris paribus preferences. *Journal of Philosophical Logic*, 38:83–125, 2009.

[vdHW05]  W. van der Hoek and M. Wooldridge. On the logic of cooperation and propositional control. *Artificial Intelligence*, 164:81–119, 2005.

[vKv07]  H. van Ditmarsch, B. Kooi, and W. van der Hoek. *Dynamic Epistemic Logic*, volume 337 of *Synthese Library Series*. Springer, 2007.

[Von63]  G. H. Von Wright. *The Logic of Preference*. Edinburgh University Press, 1963.

[Wan06]  H. Wansing. Tableaux for multi-agent deliberative-STIT logic. In G. Governatori, I. Hodkinson, and Y. Venema, editors, *Advances in Modal Logic, Volume 6*, pages 503–520. King's College Publications, 2006.

[Xu98]  M. Xu. Axioms for deliberative STIT. *Journal of Philosophical Logic*, 27:505–552, 1998.

# APPENDIX

## A. PROOF OF THEOREM 1

Let $\varphi_0$ be a PECP-formula. We have equivalence between $\varphi_0$ is PECP-satisfiable and $tr(\varphi_0)$ is S5-satisfiable.

PROOF. $\boxed{\Rightarrow}$ Suppose that there exists a PECP-model $\mathcal{M} = (W, V)$ and a world $w \in W$ such that $\mathcal{M}, w \models \varphi_0$. Let $V'$ be the valuation $V$ modified such that $p_\varphi$ is true in exactly all worlds $u$ such that $\mathcal{M}, u \models \varphi$. Let $\mathcal{M}'$ be the S5-model defined as $(W, V')$. A standard induction provides that $\mathcal{M}', w \models tr(\varphi_0)$. More precisely, let us prove by induction that for all $\varphi \in SF(\varphi_0)$, we have $\mathcal{M}, u \models \varphi$ iff $\mathcal{M}', u \models tr_1(\varphi)$ for all $u \in W$.

- Propositional case: for all atomic propositions $p$, we have $\mathcal{M}, u \models p$ iff $u \in V(p)$ iff $u \in V'(p)$ iff $\mathcal{M}', u \models tr_1(p)$.
- Negation: $\mathcal{M}, u \models \neg\varphi$ iff $\mathcal{M}, u \not\models \varphi$ iff $\mathcal{M}', u \not\models \varphi$ iff $\mathcal{M}', u \models \neg\varphi$.
- Conjunction: $\mathcal{M}, u \models \varphi \wedge \psi$ iff $\mathcal{M}, u \models \varphi$ and $\mathcal{M}, u \models \psi$ iff $\mathcal{M}', u \models tr_1(\varphi)$ and $\mathcal{M}', u \models tr_1\psi)$ iff $\mathcal{M}, u \models tr_1(\varphi \wedge \psi)$.
- Case of a formula of the form $[X]\varphi$:

  $\mathcal{M}, u \models [X]\varphi$

  iff for all $v \in W$, $u \sim_X^V v$ implies $\mathcal{M}, v \models \varphi$

  iff for all $v \in W$, $u \sim_X^V v$ implies $\mathcal{M}', v \models p_\varphi$

  (by construction of $V'$)

  iff $\mathcal{M}', u \models tr_1([X])\varphi$

By construction of $V'$, we have $\mathcal{M}', w \models \bigwedge_{\varphi \in SF(\varphi_0)} \Box(p_\varphi \leftrightarrow tr_1(\varphi))$. As $\mathcal{M}, w \models \varphi_0$ we have $\mathcal{M}, w \models tr_1(\varphi_0)$ thus $\mathcal{M}, w \models p_{\varphi_0}$ by construction of $V'$. As a result, $\mathcal{M}, w \models tr(\varphi_0)$.

$\boxed{\Leftarrow}$ Suppose that there exists a S5 model $\mathcal{M}' = (W, V)$ and a world $w \in W$ such that $\mathcal{M}', w \models tr(\varphi_0)$. We define the relations $\sim_X$ where $X \subseteq \mathbf{P}$ as in the Definition 1. Let $\mathcal{M}$ be the PECP-model equal to $(W, V)$. A standard induction provides that $\mathcal{M}, w \models \varphi_0$. More precisely, let us prove by induction that for all $\varphi \in SF(\varphi_0)$, we have $\mathcal{M}, u \models \varphi$ iff $\mathcal{M}', u \models tr_1(\varphi)$ for all $u \in W$.

- Propositional case: for all atomic propositions $p$, we have $\mathcal{M}, u \models p$ iff $u \in V(p)$ iff $u \in V'(p)$ iff $\mathcal{M}', u \models tr_1(p)$.
- Negation: $\mathcal{M}, u \models \neg\varphi$ iff $\mathcal{M}, u \not\models \varphi$ iff $\mathcal{M}', u \not\models \varphi$ iff $\mathcal{M}', u \models \neg\varphi$.
- Conjunction: $\mathcal{M}, u \models \varphi \wedge \psi$ iff $\mathcal{M}, u \models \varphi$ and $\mathcal{M}, u \models \psi$ iff $\mathcal{M}', u \models tr_1(\varphi)$ and $\mathcal{M}', u \models tr_1\psi)$ iff $\mathcal{M}, u \models tr_1(\varphi \wedge \psi)$.
- Case of a formula of the form $[X]\varphi$:

  $\mathcal{M}, u \models [X]\varphi$

  iff for all $v \in W$, $u \sim_X^V v$ implies $\mathcal{M}, v \models \varphi$

  iff for all $v \in W$, $u \sim_X^V v$ implies $\mathcal{M}', v \models tr_1(\varphi)$

  (by induction)

  iff for all $v \in W$, $u \sim_X^V v$ implies $\mathcal{M}', v \models p_\varphi$

  (because, as $\mathcal{M}', w \models tr(\varphi_0)$ we have that

  for all $v \in W$, $\mathcal{M}', v \models (p_\varphi \leftrightarrow tr_1(\varphi))$)

  iff $\mathcal{M}', u \models tr_1([X])\varphi$

As $\mathcal{M}', w \models tr(\varphi_0)$, we have that $\mathcal{M}', w \models (p_{\varphi_0} \leftrightarrow tr_1(\varphi_0))$ and $\mathcal{M}', w \models p_{\varphi_0}$. Thus, $\mathcal{M}', w \models tr_1(\varphi_0)$. Hence $\mathcal{M}, w \models \varphi_0$. □

## B. PROOF OF THEOREM 2

Let us consider a group STIT formula $\varphi$. Let $m$ be an integer. Then the following items are equivalent:

1. $\varphi$ is satisfiable in a model where each agent has at most $2^m$ actions;
2. $\varphi$ is satisfiable in a model where each agent has exactly $2^m$ actions;
3. $GRID_m \wedge tr_2(\varphi)$ is PECP-satisfiable.

PROOF. $\boxed{1 \Rightarrow 2}$ Let $\mathcal{M}^0 = (W^0, \{R_J^0\}_{J \subseteq AGT}, V^0)$ be a STIT-model with at most $2^m$ actions per agent and $w \in W^0$ such that $\mathcal{M}^0, w \models \varphi$. We construct a sequence of models $\mathcal{M}^j = (W^j, \{R_J^j\}_{J \subseteq AGT}, V^j)$ such that all agents $j' \in \{1, \ldots, j\}$ have exactly $2^m$ actions in $\mathcal{M}_j$ and such that $\mathcal{M}^j$ is bisimilar to $\mathcal{M}^{j-1}$. We construct $\mathcal{M}^j$ from $\mathcal{M}^{j-1}$ as follows. Let $R_{\{j\}}^{j-1}(w_1), \ldots, R_{\{j\}}^{j-1}(w_k)$ be an enumeration of $R_{\{j\}}^{j-1}$- classes (that is, actions for agents $j$), where $k \leq 2^m$. Let $(Copy_\ell)_{\ell \in \{k+1, \ldots, 2^m\}}$ be a family of disjoint copies of $R_{\{j\}}^{j-1}(w_1)$. We write $u\mathbf{C}v$ to say that $u = v$ or $v$ is a copy of $u$ or $u$ is a copy of $v$. The model $\mathcal{M}^j = (W^j, \{R_J^j\}_{J \subseteq AGT}, V^j)$ is defined as follows:

- $W^j = W^{j-1} \cup \bigcup_{\ell \in \{k+1, \ldots, 2^m\}} Copy_\ell$;
- $R_{\{j\}}^j = R_{\{j\}}^{j-1} \cup \bigcup_{\ell \in \{k+1, \ldots, 2^m\}} \{(u, v) \mid u, v \in Copy_\ell\}$
- $R_{\{j'\}}^j = \mathbf{C} \circ R_{\{j'\}}^{j-1} \circ \mathbf{C}$ for all $j' \neq j$;
- $V^j(p) = \{v \in W^j \mid v\mathbf{C}u \text{ and } u \in V^{j-1}(p)\}$.

This construction makes that $\mathcal{M}^j$ and $\mathcal{M}^{j-1}$ are bisimilar and by induction we have that all agents $j' \in \{1, \ldots, j\}$ have exactly $2^m$ actions in $\mathcal{M}_j$. Finally, we have $\mathcal{M}^n, w \models \varphi$ and each agent has exactly $2^m$ actions in $\mathcal{M}^n$.

$\boxed{2 \Rightarrow 3}$ Let us consider a STIT model $\mathcal{M} = (W, \{R_J\}_{J \subseteq AGT}, V)$ in which each agent has exactly $2^m$ actions. Let $w \in W$ be such that $\mathcal{M}, w \models \varphi$. For all $j \in AGT$, let $R_{\{j\}}(w_{j,1}), \ldots, R_{\{j\}}(w_{j,2^m})$ be an enumeration of all $R_{\{j\}}$-classes in $\mathcal{M}$. Let us extend $V$ such that in all worlds of $R_{\{j\}}(w_{j,i})$ the valuations of the atomic propositions in $\mathfrak{R}^j$ correspond to the binary digits in the binary representation of $i$. For all $d \in \{1, \ldots, m\}$:

- $V(\mathfrak{rep}_d^j) = \bigcup_{i=1..2^m \mid \text{ the } d^{th} \text{ digit of } i \text{ is } 1} R_{\{j\}}(w_{j,i})$

Independence of agents in $\mathcal{M}$ ensures that $\mathcal{M}, w \models GRID_m$. We prove that $\mathcal{M}, u \models tr_2(\psi)$ iff $\mathcal{M}, u \models \psi$ by induction over all subformulae $\psi$ of $\varphi$.

$\boxed{3 \Rightarrow 1}$ Let $\mathcal{M} = (W, V)$ be a PECP-model and $w \in W$ such that $\mathcal{M}, w \models GRID_m \wedge tr_2(\varphi)$. We define $R_J = \sim_{\bigcup_{j \in J} \mathfrak{R}^j}$. The resulting Kripke-model $\mathcal{M}' = (W, \{R_J\}_{J \subseteq AGT}, V)$ is a STIT-model where each agent has exactly $2^m$ actions. In particular, it satisfies the independence of agents because $\mathcal{M}, w \models GRID_m$. We prove that $\mathcal{M}, u \models tr_2(\psi)$ iff $\mathcal{M}', u \models \psi$ by induction over all subformulae $\psi$ of $\varphi$. □

## C. PROOF OF THEOREM 4

The PECP-nested satisfiability problem is PSPACE-hard.

PROOF. We reduce the satisfiability problem of STIT-formulae where coalitions are taken from a linear set of coalitions, which is PSPACE-complete [Sch12] to the PECP-nested satisfiability problem: we use the translation $tr_2$ of Subsection 3.1. Let $\varphi$ be a STIT-formula. We have $\varphi$ is STIT-satisfiable iff $tr_2(\varphi)$ is PECP-satisfiable.

$\boxed{\Rightarrow}$ As it stated in [Sch12], the STIT where coalitions are taken from a linear set of coalitions has the exponential model property. So the result of Theorem 2 is true. Hence if $\varphi$ is STIT-satisfiable then $GRID_m \wedge tr_2(\varphi)$ is PECP-satisfiable (where $m$ is the length of $\varphi$). Hence $tr_2(\varphi)$ is PECP-satisfiable.

$\boxed{\Leftarrow}$ Suppose that there exists a PECP-model $\mathcal{M} = (W, V)$ and $w \in W$ such that $\mathcal{M}, w \models tr_2(\varphi)$. We define $R_J = \sim_{\bigcup_{j \in J}}$. Then the STIT model $\mathcal{M}' = (W, (R_J)_{J \in \varphi}, V)$ is such that $\mathcal{M}', w \models \varphi$. Remark that we do not need to specify all the relations $R_J$ for all $J$. As long as $R_J$ is specified for all coalitions $J$ that appear in $\varphi$ and that $R_J \subseteq R_{J'}$ if $J' \subseteq J$, we can extend the Kripke model $\mathcal{M}'$ to a completely specified STIT-model also satisfying $\varphi$.[6] □

## D. PROOF OF THEOREM 5

The PECP-nested satisfiability problem is in PSPACE.

PROOF. We reduce the PECP-nested satisfiability problem to the satisfiability problem of STIT where coalitions are taken from a linear set of coalitions. We define the set $A_X = \{j_p \text{ such that } p \in X\}$ where $j_p$ is a fresh agent corresponding to the atomic proposition $p$. Let us define the following translation:

- $tr_4(p) = p$;
- $tr_4(\neg\varphi) = \neg tr_4(\varphi)$;
- $tr_4(\varphi \wedge \psi) = tr_4(\varphi) \wedge tr_4(\psi)$;
- $tr_4([X]\varphi) = [A_X : stit]tr_4(\varphi)$.

Let us consider a fixed PECP-formula $\varphi$. We recall that a signature $X$ appears in $\varphi$ if there exists a formula $\psi$ such that $\langle X \rangle \psi \in SF(\varphi)$. We have also to define the following formula

$$CONTROL = [\emptyset : stit]$$
$$\bigwedge_{X \text{ appearing in } \varphi}$$
$$\bigwedge_{p \in X}(p \leftrightarrow [A_X : stit]p) \wedge$$
$$(\neg p \leftrightarrow [A_X : stit]\neg p).$$

$tr_4(\varphi) \wedge CONTROL$ is a STIT-formula which is computable in polynomial time and which satisfies the condition of nesting over groups (i.e., for any two operators $[J : stit]$ and $[J' : stit]$ occurring in the formula either $J \subseteq J'$ or $J' \subseteq J$). We also have that $\varphi$ is PECP-satisfiable iff $tr_4(\varphi) \wedge CONTROL$ is satisfiable in a STIT-model.

$\boxed{\Rightarrow}$ Suppose that there exists an PECP-model $\mathcal{M} = (W, V)$ and $w \in W$ such that $\mathcal{M}, w \models \varphi$. We define $R_{A_X} = \sim_X$. Then the STIT model $\mathcal{M}' = (W, (R_{A_X})_{X \in \varphi}, V)$ is such that $\mathcal{M}', w \models tr_4(\varphi) \wedge CONTROL$. Remark that we do not need to specify all the relations $R_J$ for all $J$. As long as $R_J$ is specified for all coalitions $J$ that appear in $tr_4(\varphi) \wedge CONTROL$ and that $R_J \subseteq R_{J'}$ if $J' \subseteq J$, we can extend the Kripke

---

[6]See [Sch12] for more details about this construction.

model $\mathcal{M}'$ to a completely specified STIT-model also satisfying $tr_4(\varphi) \wedge CONTROL$.[7]

$\boxed{\Leftarrow}$ Suppose that there exists a STIT-model $\mathcal{M}' = (W, (R_{A_X})_{X \in \varphi}, V)$ and a world $w \in W$ such that $\mathcal{M}', w \models tr_4(\varphi) \wedge CONTROL$. As $\mathcal{M}', w \models CONTROL$, we have $\sim_X = R_{A_X}$. This is the reason why if we define $\mathcal{M} = (W, \{\sim_X\}_{X \in 2^\mathbf{P}}, V)$. Consequently, we have $\mathcal{M}, w \models \varphi$. $\square$

# E. PROOF OF THEOREM 6

Let $m = |\mathbf{P}|$. Then, a CL−PC formula $\varphi$ is CL−PC satisfiable if and only if $(EXC^+ \wedge EXC^- \wedge COMPL \wedge GRID^*) \wedge tr_3(\varphi)$ is satisfiable in a STIT model where each agent has at most $2^m$ actions.

PROOF. Let us suppose $|\mathbf{P}| = m$.

$\boxed{\Rightarrow}$ Let $\mathcal{M}^* = (\mathbf{P}_1, \ldots, \mathbf{P}_n, X^*)$ be a CL−PC model such that $\mathcal{M}^* \models \varphi$, where $\mathbf{P}_1, \ldots, \mathbf{P}_n$ is a partition of $\mathbf{P}$ among the agents in $AGT$. We build the STIT model $\mathcal{M} = (W, \{R_J\}_{J \subseteq AGT}, V)$ as follows:

- $W = \{X : X \subseteq \mathbf{P}\}$,
- for all $J \subseteq AGT$ and for all $X, X' \in W$, $(X, X') \in R'_J$ if and only if $X_J = X'_J$,
- for all $p \in \mathbf{P}$ and for all $X \in W$, $X \in V(p)$ if and only if $p \in X$,

where for any $X \subseteq \mathbf{P}$ and for any $J \subseteq AGT$, $X_J = X \cap \mathbf{P}_J$ (with $\mathbf{P}_J = \bigcup_{i \in J} \mathbf{P}_i$). The size of $\mathcal{M}$ is $2^m$. It follows that the number of $R_{AGT}$-equivalence classes (*alias* joint actions) is equal or lower than $2^m$. Consequently, the number of actions for every agent is bounded by $2^m$.

It is straightforward to prove that for all $X \in W$ we have $\mathcal{M}, X \models EXC^+ \wedge EXC^- \wedge COMPL \wedge GRID^*$. Moreover, by induction on the structure of $\varphi$, we prove that $\mathcal{M}, X^* \models tr_3(\varphi)$. The only interesting case is $\varphi = \Diamond_J \psi$:

$\mathcal{M}^* \models \Diamond_J \psi$ iff there exists $X_J \subseteq \mathbf{P}_J$ s.t. $\mathcal{M}^* \bigoplus X_J \models \psi$

iff there exists $X_J \subseteq \mathbf{P}_J$ s.t.

$\mathcal{M}, X_J \cup X^*_{AGT \setminus J} \models tr_3(\psi)$ (by I.H.)

iff $\mathcal{M}, X^* \models \langle AGT \setminus J : stit \rangle tr_3(\psi)$

$\boxed{\Leftarrow}$ Let $\mathcal{M} = (W, \{R_J\}_{J \subseteq AGT}, V)$ be a STIT model where the number of actions for every agent is bounded by $2^m$ and $w_0 \in W$ such that $\mathcal{M}, w_0 \models (EXC^+ \wedge EXC^- \wedge COMPL \wedge GRID^*) \wedge tr_3(\varphi)$.

For any $i \in AGT$, let

$Ctrl_i = \{p \in \mathbf{P} : \forall v \in W, \begin{matrix} \mathcal{M}, v \models [\{i\} : stit]p \text{ or} \\ \mathcal{M}, v \models [\{i\} : stit]\neg p \end{matrix} \}$

be the set of atoms in $\mathbf{P}$ controlled by agent $i$. For any $J \subseteq AGT$, let $Ctrl_J = \bigcup_{i \in J} Ctrl_i$.

LEMMA 1. *For all $J \subseteq AGT$, $X \subseteq \mathbf{P}$, $\pi_X \subseteq X$ and $w \in W$ we have:*

(i) *if $Ctrl_J = X$ then $Ctrl_{AGT \setminus J} = \mathbf{P} \setminus X$,*

(ii) *if $\mathcal{M}, w \models \bigwedge_{p \in \pi_X^+} p \wedge \bigwedge_{p \in \pi_X^-} \neg p$ and $Ctrl_J = X$ then, for all $v \in R_J(w)$, we have $\mathcal{M}, v \models \bigwedge_{p \in \pi_X^+} p \wedge \bigwedge_{p \in \pi_X^-} \neg p$,*

(iii) *if $Ctrl_J = X$ then, for all $\pi'_{\mathbf{P} \setminus X} \subseteq \mathbf{P} \setminus X$, there exists $v \in R_J(w)$ such that $\mathcal{M}, v \models \bigwedge_{p \in \pi'^+_{\mathbf{P} \setminus X}} p \wedge \bigwedge_{p \in \pi'^-_{\mathbf{P} \setminus X}} \neg p$.*

---

[7]Again see [Sch12] for more details about this construction.

where for any $X \subseteq \mathbf{P}$ and for any $\pi_X \subseteq X$, $\pi_X^+ = \pi_X$ and $\pi_X^- = X \setminus \pi_X$.

PROOF. $\boxed{(i)}$ Let us suppose that $p \notin Ctrl_J$. We are going to prove that $p \in Ctrl_{AGT \setminus J}$. From $p \notin Ctrl_J$ it follows that for all $w \in W$ we have $\mathcal{M}, v \models \neg p$ for some $v \in R_J(w)$. This implies that for all $i \in J$ and for all $w \in W$ we have $\mathcal{M}, w \models \neg[\{i\} : stit]p \wedge \neg[\{i\} : stit]\neg p$. From $\mathcal{M}, w_0 \models COMPL$ it follows that there is $i \in AGT \setminus J$ such that $\mathcal{M}, w \models [\{i\} : stit]p \vee [\{i\} : stit]\neg p$ for all $w \in W$. The latter implies that $p \in Ctrl_{AGT \setminus J}$. The other direction (i.e., $p \in Ctrl_J$ implies $p \notin Ctrl_{AGT \setminus J}$) follows from $\mathcal{M}, w_0 \models EXC^+ \wedge EXC^-$.

$\boxed{(ii)}$ Let us suppose that $\mathcal{M}, w \models \bigwedge_{p \in \pi_X^+} p \wedge \bigwedge_{p \in \pi_X^-} \neg p$ and $Ctrl_J = X$. By the fact that relations $R_J$ are reflexive, it follows that, for all $p \in \pi_X^+$, there exists $i \in J$ such that $\mathcal{M}, w \models [\{i\} : stit]p$ and for all $p \in \pi_X^-$ there exists $i \in J$ such that $\mathcal{M}, v \models [\{i\} : stit]\neg p$. From the latter it follows that for all $p \in \pi_X^+$ we have $\mathcal{M}, w \models [J : stit]p$ and for all $p \in \pi_X^-$ we have $\mathcal{M}, v \models [J : stit]\neg p$. Therefore, for all $v \in R_J(w)$, we have $\mathcal{M}, v \models \bigwedge_{p \in \pi_X^+} p \wedge \bigwedge_{p \in \pi_X^-} \neg p$.

$\boxed{(iii)}$ Let us suppose that $Ctrl_J = X$ and let us consider an arbitrary $\pi'_{\mathbf{P} \setminus X} \subseteq \mathbf{P} \setminus X$ and $w \in W$. From $\mathcal{M}, w_0 \models GRID^*$ it follows that there exists $v \in W$ such that $\mathcal{M}, v \models \bigwedge_{p \in \pi'^+_{\mathbf{P} \setminus X}} p \wedge \bigwedge_{p \in \pi'^-_{\mathbf{P} \setminus X}} \neg p$. By item (ii), the latter implies that there exists $v \in W$ such that $\mathcal{M}, v \models [AGT \setminus J : stit](\bigwedge_{p \in \pi'^+_{\mathbf{P} \setminus X}} p \wedge \bigwedge_{p \in \pi'^-_{\mathbf{P} \setminus X}} \neg p)$. From the constraint of *independence of agents* it follows that there exists $v \in R_J(w)$ such that $\mathcal{M}, v \models \bigwedge_{p \in \pi'^+_{\mathbf{P} \setminus X}} p \wedge \bigwedge_{p \in \pi'^-_{\mathbf{P} \setminus X}} \neg p$. $\square$

We transform the STIT model $\mathcal{M}$ in a CL−PC model $\mathcal{M}^* = (\mathbf{P}_1, \ldots, \mathbf{P}_n, X^*)$ as follows:

- for all $p \in \mathbf{P}$, $p \in X^*$ if and only if $w_0 \in V(p)$,
- for all $p \in \mathbf{P}$ and for all $i \in AGT$, $p \in \mathbf{P}_i$ if and only if $p \in Ctrl_i$.

By the item (i) of Lemma 1 it is easy to check that $\mathcal{M}^*$ is indeed a CL−PC model. In particular, $\mathbf{P}_1, \ldots, \mathbf{P}_n$ is a partition of $\mathbf{P}$ among the agents in $AGT$.

By induction on the structure of $\varphi$ and by using Lemma 1 it is straightforward to prove that $\mathcal{M}^* \models \varphi$. The only interesting case is $\varphi = \Diamond_J \psi$:

$\mathcal{M}, w_0 \models \langle AGT \setminus J : stit \rangle tr_3(\psi)$

iff $\mathcal{M}, v \models tr_3(\psi)$ for some $v \in R_{AGT \setminus J}(w_0)$

iff there exists $X_J \subseteq \mathbf{P}_J$ s.t.

$(\mathbf{P}_1, \ldots, \mathbf{P}_n, X_J \cup X^*_{AGT \setminus J}) \models \psi$

(by I.H., and items (ii) and (iii)

of Lemma 1)

iff $\mathcal{M}^* \models \Diamond_J \psi$

This completes the proof.